\documentclass[12pt,usenatbib]{emulateapj}
\slugcomment{paper draft, \today}
\usepackage{natbib}
\usepackage{epsfig}

\begin{document}

\title{The Space Density Evolution of Wet and Dry Mergers in the Canada-France-Hawaii Telescope Legacy Survey }

\author{Richard C. Y. Chou}

\affil{Department of Astronomy and Astrophysics, University of Toronto, 50 St. George Street, Toronto, ON M5S 3H4}
\email{chou@astro.utoronto.ca}

\author{Carrie R. Bridge}
\affil{California Institute of Technology, 1200 East California Blvd. 91125 }
\email{bridge@astro.caltech.edu}

\author{Roberto G. Abraham}
\affil{Department of Astronomy and Astrophysics, University of Toronto, 50 St. George Street, Toronto, ON M5S 3H4}
\email{abraham@astro.utoronto.ca}

\begin{abstract}

We analyze 1298 merging galaxies with redshifts up to 
$z=0.7$ from the Canada-France-Hawaii Telescope Legacy Survey, taken from the
catalog presented in Bridge et al. (2010).
By analyzing the internal colors of these systems, we show that
so-called wet and dry mergers evolve in different senses, and
quantify the space densities of these systems.
The local space density of wet mergers is essentially identical to the local
space density of dry mergers. The
evolution in the total merger rate is
modest out to $z\sim0.7$, although the wet and dry
populations have different
evolutionary trends. At higher redshifts 
dry mergers make a smaller contribution to the
total merging galaxy population, but this is
offset by a roughly equivalent increase in the contribution
from wet mergers.
By comparing the mass density function of early-type galaxies
to the corresponding mass density function for merging systems,
we show that not all the major mergers with the highest masses (M$_{\rm stellar} > 10^{11}$M$_{\odot}$) will end up with the most massive
early-type galaxies, unless the merging timescale is dramatically longer
than that usually assumed. On the other hand, the usually-assumed merging
timescale of $\sim0.5-1$~Gyr is quite consistent with the data if we
suppose that only less massive
early-type galaxies form via mergers. 
Since low-intermediate mass ellipticals
are 10--100 times more common than their most massive counterparts, the hierarchical
explanation for the origin of early-type galaxies may be correct for the vast majority of
early-types, even if incorrect for the most massive ones. 
\vspace{0.5cm}  
\end{abstract}

\section{INTRODUCTION}
\label{intro}

Hierarchical structure formation models suggest that galaxy mergers play an important role in galaxy mass assembly, but quantifying that role has remained challenging.  The galaxy merger rate is generally parametrized by a power law of the form $(1+z)^m$, and the value of the exponent of this parametric form has been used to place constraints on how much mass is assembled via major galaxy mergers.  
Large variations in $m$ are found in the literature, ranging from $m\sim0$ to $m\sim4$ \citep{bundy04,bundy09,conselice03,gw08,lin04,bridge07,lin08,lotz08,jog09}.
A recent study by \citet{bridge10} analyzed these published merger rates and concluded that, overall,
there is a general agreement that the merger rate at intermediate redshifts ($0.2<z<1.2$) does evolve, although
the constraints on $m$ remain fairly mild.  Bridge et al. (2010) rule out $m<1.5$ (i.e. flat or mild evolution)
and suggest  that the wide range of $m$ reported in the literature is due to a combination of factors,
including variation in the redshift ranges being probed, small sample sizes in some
of the surveys, and cosmic variance.  

The traditional approach used to explore the merger history of galaxies has been to estimate the fraction of merging galaxies relative to the total galaxy population.  This approach has the benefit of simplicity, but it is arguable that a more physically interesting quantity is  the evolving space density of mergers, rather than the merger fraction.
Space densities are absolute measurements rather than relative measurements, and in that sense can stand on their own.
Furthermore, space densities can be corrected for luminosity biases and other sources of incompleteness in a straightforward manner by using the standard $1/V_{\rm max}$ formalism \citep{felten77,schmidt68}. Therefore our main aim in the present paper
is to chart the evolving space density of mergers. Similar work has been undertaken by \cite{lin08} and \cite{bundy09}, although
these papers used pair counts to select merging galaxies, while 
our approach is based on morphological selection. Our analysis is thus quite complementary
to \cite{lin08} and \cite{bundy09}.

An important subsidiary goal of the present paper is to chart the differential merging history of color-selected sub-classes of 
merging galaxies.
In recent years a host of observations have shown the evolutionary histories of galaxies in the so-called `red sequence' and and `blue cloud' are different  \citep{bundy09,deravel09,lin08,willmer06}. This has led to the idea that it is important
to distinguish between mergers that result in significant star-formation (`wet mergers') and those which merely
re-organize existing stellar populations (`dry mergers').
Wet mergers are typically associated with gas-rich systems
that trigger star formation \citep{barton00,bridge07,lin07,overzier08,matteo08,zaur10}, cause quasar activity \citep{hopkins06} and turn disk galaxies into elliptical galaxies \citep{tt72}.  On the other hand, dry mergers \citep{bell06,dokkum05} may
be responsible for the assembly of massive (M$_{stellar} \gtrsim 10^{11}$ M$_{\odot}$) red galaxies which are observed in surprisingly high abundance at $z\sim1$.  However, the importance of these wet and dry mergers in the formation of red sequence galaxies is still not clear \citep{bell07,brown07,bundy09,faber07,lin08,scarlata07}, and it is of interest to determine if wet and dry merging
systems exhibit similar evolutionary trends as a function of cosmic epoch. 

A plan for this paper follows. In Section 2 we describe the observations, galaxy properties and merger identifications for our sample. This section is essentially a brief recapitulation, presented for the convenience of the reader, of the comprehensive description of the data set given in Bridge et al. (2010).  In Section 3 we define our methodology for defining `wet' and `dry' mergers on the basis of both resolved colors and integrated colors. Section 4 is the heart of the paper, where we present the merger fraction, number density and stellar mass density for dry and wet mergers.  
Since recent work shows that the formation rate of massive elliptical galaxies through dry mergers is dependent on stellar masses \citep{bell07,bundy09,ks09}, in this section
we also investigate the evolving space density of mass-segregated samples of wet and dry mergers. 
Our results are discussed in Section 5, and our conclusions presented in Section 6

Throughout this paper, we
adopt a concordance cosmology with \hbox{$H_0$=70 km s$^{-1}$ Mpc$^{-1}$},
$\Omega_M=0.3$, and $\Omega_\Lambda=0.7$. All photometric magnitudes are given
in the AB system.

\begin{figure}[htb]
\includegraphics[width=8.5 cm,angle=0]{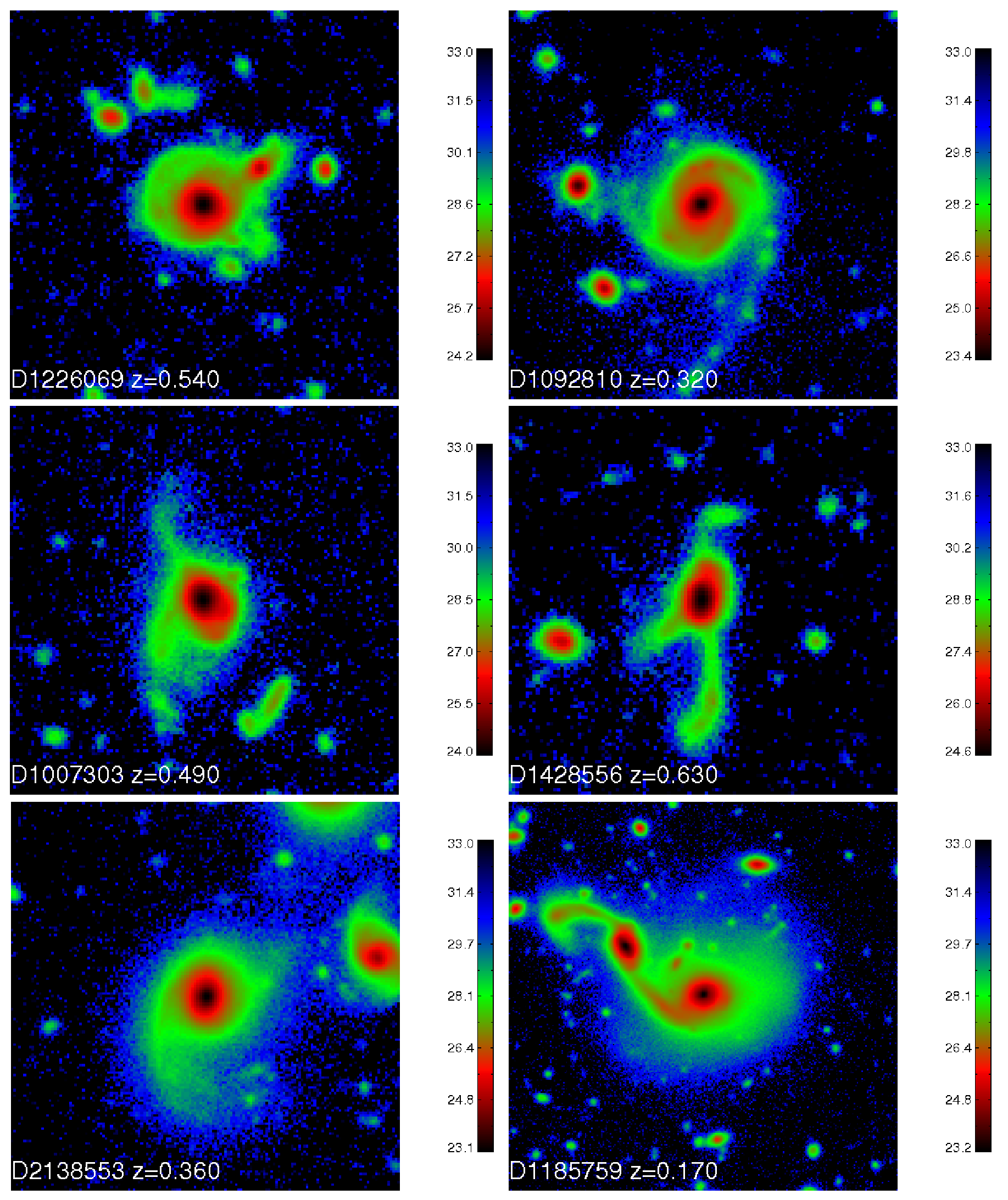}
\caption{A montage of $i'$-band image of six typical galaxy mergers.  Each stamp has the dimension of $100 \times 100$ kpc.  The catalog name and the photometric redshift of the merger is shown in the bottom left corner of each stamp. }
\label{montage_i}
\end{figure}

\section{OBSERVATIONS}
\label{obs}

As has already been noted, a detailed description of the selection strategy for (and basic properties of) the galaxies analyzed in the present paper has already been presented in Bridge et al. (2010). The reader is referred to that paper for details beyond the outline
 presented here.

\subsection{Data}
The data in this paper come from two of the Canada-France-Hawaii Telescope Legacy Survey (CFHTLS) deep survey fields. These fields (denoted D1 and D2) together cover an area of 2 square degrees. The CFHTLS deep survey has high-quality broad-band photometry in five bands ($u^*$, $g'$, $r'$,$i'$,$z'$) and the depth of the survey ranges from 26.0 ($z'$) to 27.8 ($g'$).  The optical images used to derive galaxy properties and morphological classifications were stacked by the  {\tt Elixir} image processing pipeline \citep{mc04} to produce deep optical stacks with precise astrometric solutions.  The typical seeing for the final stacks is 0.7''-0.8'' in the $i'$-band.  The source extraction and photometry were performed on each field using SExtractor \citep{ba96} in dual image mode.  The source detection was performed in the $i'$ filter ($i' \sim 26.3$).  A bad pixel mask was applied to the image prior to running the program to avoid noisy or contaminated regions caused by spikes or halos of bright stars.  The total area masked is less than 10\% for each field.

\begin{figure}[htb]
\includegraphics[width=7.5cm,angle=0]{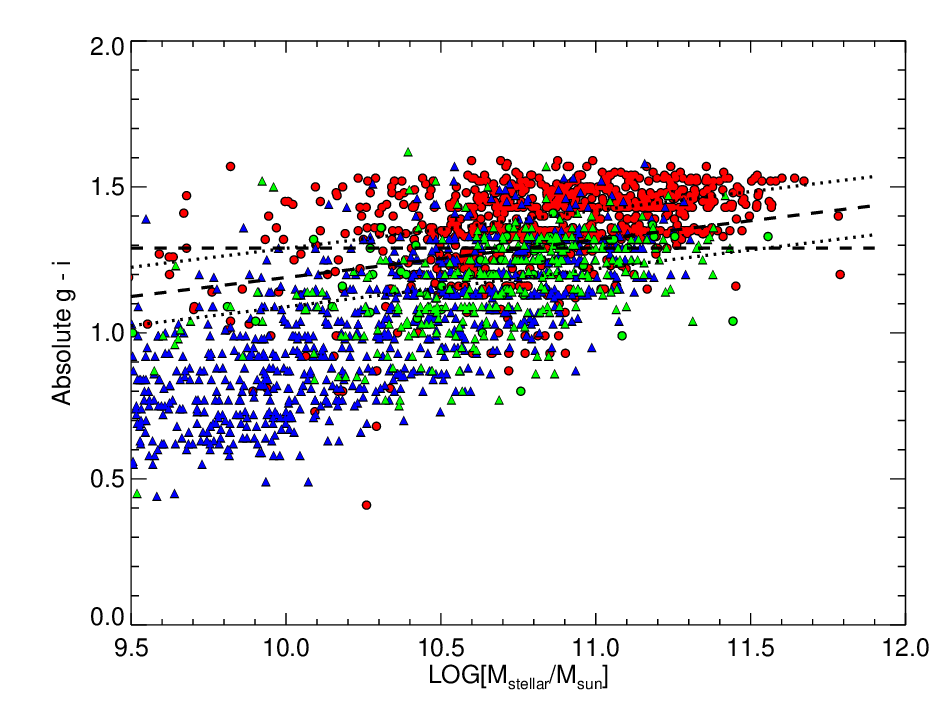}
\caption{
Colors of $\sim 2200$ field galaxies in the CFHTLS survey.  The red filled circles and blue triangles represent the visually classified elliptical galaxies and spiral galaxies, respectively.  The dashed line with a slope of 0.13 is derived from the red sequence fitting of 115 visually-classified elliptical galaxies with spectroscopic redshifts from the COSMOS; two equally spaced parallel dotted lines are used for merger classification test (described in the text).  The horizontal dashed line represents rest frame color $g'-i'=1.29$ mag, and is also used for the same test.  The green circles and triangles represent elliptical galaxies and spiral galaxies with MIPS 24$ \mu m $ detections.  As described in the text, we assign the MIPS-detected galaxies with $g'-i'$ color greater than the color bimodality line to the blue population.  The results of merger classification test is shown in Figure \ref{merger_fraction}, see text for more information.}
\label{color-bimo}
\end{figure}

\subsection{Merger identifications}
\label{sec:merger_id}
Merging galaxies were selected visually, with multiple cross-checks on the visual classifications, and using simulations to characterize detection thresholds for features that are signatures of mergers.  Visual classifications are labor intensive and somewhat subjective, but remain the best method presently available if accuracy is the ultimate goal.  Interacting galaxies are defined as
systems with a tidal tail or bridge. All galaxies down to an $i'_{vega} \leq 22.9$ mag ($\sim$ 27,000) were inspected resulting in a final 
sample of $1298$ merging galaxies.  The merger identification rate for galaxies with $i'_{vega} \leq 21.9$~mag is estimated to be 
$> 90\%$, and drops rather quickly at fainter magnitudes. Therefore in the present paper the $i'_{vega} = 21.9$ mag is used as the limit for merger identification.
The redshift completeness limit was estimated by artificially redshifting low-redshift galaxies. For this completeness test, a group of galaxy mergers with redshift ranges from $z = 0.3 $ to $ z = 0.45$ and M$_g$ $\leq -21.0$~mag were selected.  They were artificially redshifted to higher redshifts after accounting for the k-correction,  change in angular size and surface brightness dimming.  After this step the merger identification was conducted once again and found to be acceptably high out to  $z \sim 0.7$, where $\sim 85$\% of the redshifted mergers could be still be identified as mergers, and the false positive contamination fraction remains minimal. This redshift limit is particularly important because we utilize a $1/V_{\rm max}$ weighting to correct the Malmquist bias and to compute number densities. {\em It is important to note that $V_{\rm max}$ of mergers presented in this paper is defined as the maximum volume over which mergers can be identified as such, and not the maximum redshift at which a given galaxy's integrated magnitude remains above the detection threshold.}  As described in the next section, we calculated the $V_{\rm max}$ value from the $z_{\rm max}$ provided by the Z-Peg code which denotes the maximum redshift that the template SEDs is fainter than the observed limiting magnitude ($i'_{vega} \leq 21.9$ mag).

\subsection{Galaxy properties}

The CFHTLS survey has high quality five broad band photometry which makes the derivation of accurate photometric redshifts, ages and stellar masses possible.  The galaxy properties were derived by comparing the spectral energy distributions (SEDs) obtained from observed fluxes to a set of template SEDs.  The best-fit SEDs were determined through a standard minimum $\chi ^2$ fitting between the template SEDs and the observed fluxes.  The template SEDs were computed by the PEGASE-II galaxy evolution code \citep{fr97,lr02,leborgne04} and were integrated through the CFHT filters.  The SED fits were undertaken using the Z-Peg code \citep{bolzonella00,lr02} and details are described in \citet{bridge10}.  The photometric accuracy is determined by comparing the derived photometric redshifts to the spectroscopic redshifts in the SNLS sample \citep{howell05,bronder08}.  The accuracy of the photometric redshift down to $i′ \sim 22.5$ is $\sigma_{\Delta z}/ (1+z) = 0.04$.  The stellar mass for each merger was also estimated using the Z-Peg code by integrating the total star formation history (SFH) of the best-fit model, up to the best-fit age and subtracting off mass loss from late stages of stellar evolution.

%

\section{Classification of Wet and Dry Mergers}

We used two techniques to try to distinguish between wet and dry mergers.  The common starting point for both methods is subdivision into ÔredÕ and ÔblueÕ stellar populations on the basis of rest-frame color relative to a fiducial reference color (Van Dokkum 2005; Willmer et al. 2006). Our approach to defining this reference color uses the rest frame $g' - i'$ versus $g'$ color-magnitude diagram for $\sim 2200$ visually classified field galaxies in the CFHTLS D1 and D2 field (see Figure \ref{color-bimo}).  Red dots represent visually classified elliptical galaxies and blue dots indicate spiral galaxies.  The green dots indicate the objects with MIPS 24$\mu m$ detection (down to a flux limit of 340 $\mu$Jy).  \citet{cowie08} report that at $z<1.5$ most red galaxies with a 24$\mu m$ flux $>80 \mu \rm Jy$ fall into the blue cloud after the appropriate dust extinction is applied.  To account for the color change in dusty sources, we artificially assign the green dots with $g'-i'$ color greater than the color bimodality to the blue cloud.  In addition, color bimodality is known to be magnitude or stellar mass-dependent and usually derived from the fitting of red sequence objects \citep{dokkum00, willmer06}.  Therefore, the following stellar-mass-dependent fiducial color cut was adopted based on the red sequence fitting of 115 visually classified elliptical field galaxies with spectroscopic redshifts from Cosmological Evolution Survey (COSMOS) \citep{scoville07,lilly07}.  The fitting line is expressed by the following equation:

\begin{equation}
  (g' - i')_{\rm rest} = -0.0076+0.13\times M_{stellar} - C 
 \end{equation}
 \smallskip

\noindent The constant $C$ serves as a parameter to control the vertical position of the fitting line on the diagram.  To account for the potential classification errors caused by different slopes and $C$ values, we have explored the implications of changing the
free parameters in Equation~(1) and find that all trends reported in this paper remain robust to the specific numerical values chosen.
To show this, in several key figures in this paper we will bracket the results obtained using Equation~(1) with
curves showing the envelope obtained by using 
three values of $C$ (0, 0.1 and 0.2)  as well as using a
perfectly horizontal line set at $g' - i' =1.29$ mag to divide field galaxies into blue and red clouds\footnote{The reason to use $g' - i' =1.29$ is that we try to balance the contamination of blue (red) galaxies in red (blue) cloud.  For the $g' - i' =1.29$ color cut the contamination on both sides is 14\%. }


\subsection{Method I: integrated colors}

The first method for segregating wet mergers from dry mergers is, essentially, the simplest conceivable. We look at the total integrated
color of the merger which we obtained by summing over all pixels in the merger (or merging pair), and simply note if the
integrated color of the complete system is redder or bluer than the fiducial  color threshold. Mergers whose integrated colors are redder than the threshold are deemed `dry', and systems bluer than the threshold are deemed `wet'.  As mentioned above, the
only caveat is that we re-assign dry mergers with 24$\mu$m detections to the wet merger category on the assumption that these are blue galaxies being reddened by dust.  

This simple method is straightforward but it is not at all obvious that loss of the spatial information is an acceptable trade-off for such simplicity. For example, what if one object in a merging system is redder than the fiducial color, while the other object is bluer? We therefore decided to explore a somewhat more refined approach to `wet' vs. `dry' merger classification that retains some component of the spatial information in the images.  

\subsection{Method II: spatially resolved colors}

Our second method is based on analysis of the colors of individual pixels.
Pixels with rest-frame $g'$ - $i'$ color greater than the fiducial 
threshold are labeled as `red', and the ratio of the total flux in red pixels to the flux in all pixels is calculated.
We refer to this quantity as the `Red-to-Total ratio', $(R/T)$, given by:
\begin{equation}
(R/T) = \frac{F_{red}}{F_{total}}
\end{equation}
where $F_{red}$ indicates flux contained in red pixels and $F_{total}$ refers to the flux from the entire merger.

The idea behind $(R/T)$ is to crudely decompose a merger into individual stellar populations. A galaxy with
$(R/T)=1$ is comprised entirely of stellar populations with colors on the
red sequence, while a merger with $(R/T)=0$ is comprised entirely of stellar populations 
in the blue cloud. Of course most mergers are expected to lie somewhere in between 
these extremes. For the sake of simplicity we choose to define systems with
$(R/T)>0.5$ as being `dry', but throughout this paper we will explore the results obtained
for range of $(R/T)$ values to ensure that our conclusions are not tied to any specific value of $(R/T)$.  
Finally, we note that
in common with Method I, we re-assign `dry' mergers with MIPS 24 $\mu m$ detections into the blue cloud.

\section{RESULTS}
\label{results}

\begin{figure*}[htb]
\begin{center}
\includegraphics[width=7.5cm,angle=0]{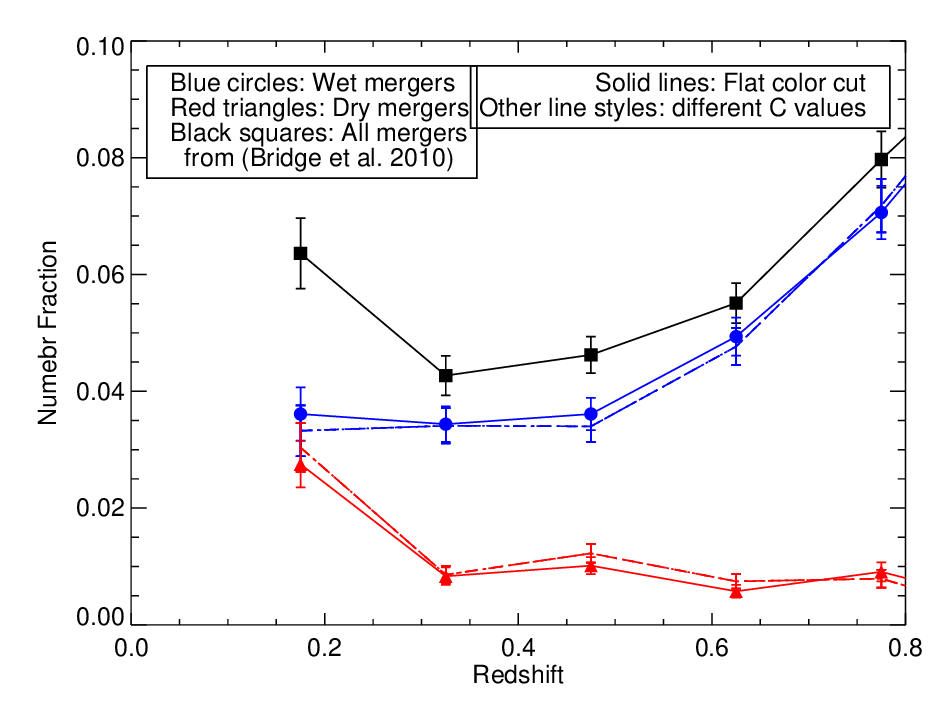}
\includegraphics[width=7.5cm,angle=0]{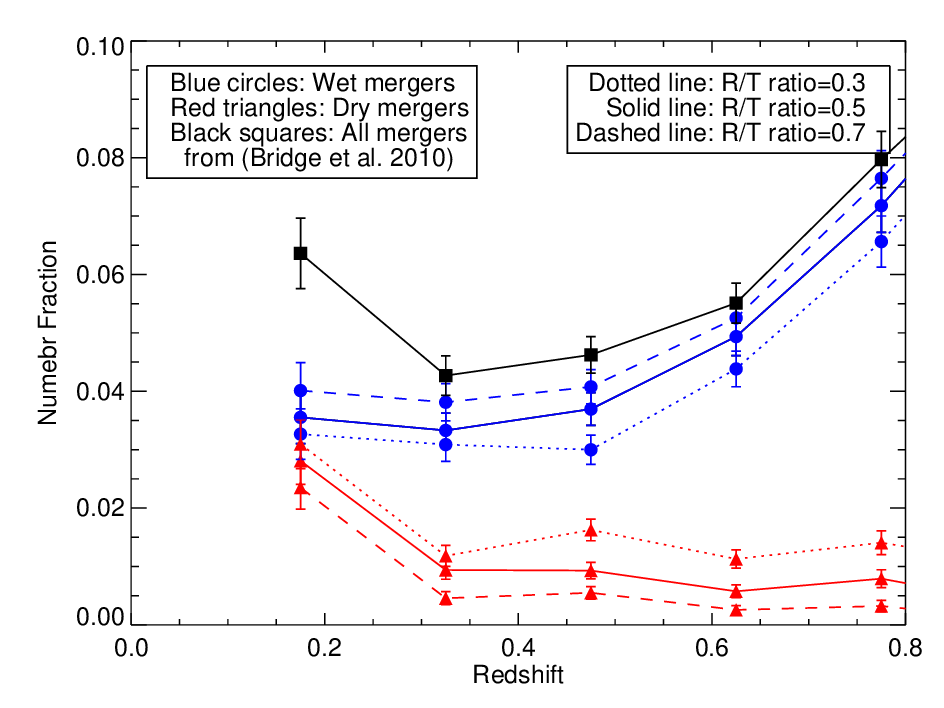}
\end{center}
\caption{
Merger fractions computed using two different methods.The left panel shows the merger fraction derived from different fiducial color cuts with integrated colors, while the right panel shows the merger fraction derived from different internal color ratios (see text for details). It is obvious that different color cuts do not affect the scientific results.  In both panels,
red curves indicate dry mergers, blue curves indicate wet mergers, and black curves indicate the total merging population.
It is clear that the increase with redshift in the wet merger fraction, and the decrease with redshift in the dry merger fraction, are robust
and do not depend on the selection method. Error bars are estimated by assuming Poisson errors. }
\label{merger_fraction}
\end{figure*}

\begin{figure*}[htb]
\begin{center}
\includegraphics[width=10cm,angle=0]{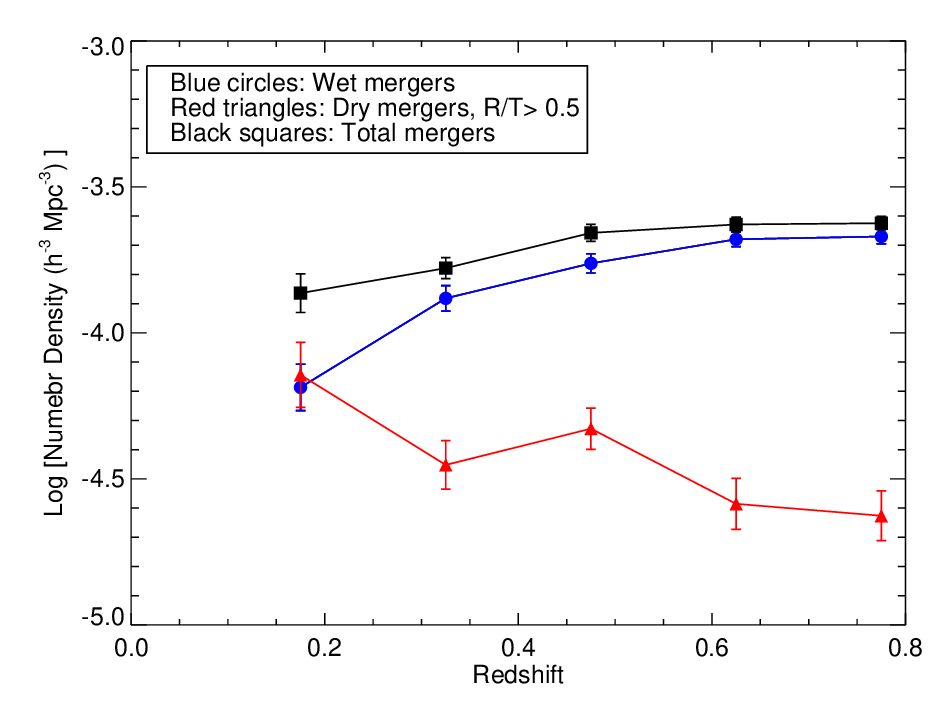}
\end{center}
\caption{Space densities of wet and dry mergers.  As in the previous figure,  blue and red curves correspond to
 wet and dry mergers, respectively, while the black curve shows the total for all mergers. 
 The growth in the total space density of mergers is modest, increasing by about a factor of two
 over the redshift range probed. The wet and dry merging galaxies show opposite trends with redshift.
 The space density of wet mergers is increasing with redshift, while that of dry mergers is 
 (perhaps) modestly decreasing.}
\label{number_density}
\end{figure*}

\subsection{Merger Fractions}
Figure  \ref{merger_fraction} shows the merger fraction for a stellar mass-limited 
sample of objects (M$_* > 10^{9.5}$ M$_{\odot}$), $f_M$, as a function of
redshift. We define this merger
fraction in the obvious way as $f_M={N_M}/N$,
where  $N_M$ is the number of mergers and $N$ is the number of galaxies in the complete
sample (once again, we refer the reader to Bridge et al. 2010 for details of the full sample). The figure shows
curves for wet mergers (blue lines), dry mergers (red lines), and all mergers (black lines). 
The left-hand panel shows the merger fraction
computed using integrated colors while the right-hand panel shows the corresponding
curves computed using resolved colors, at three different values of the red-to-total
fraction used to segregate wet and dry mergers. 
Errors were estimated assuming Poisson statistics.

A number of interesting results emerge from these figures. Firstly, a comparison of the
left and right panels shows that the general trends seen are quite independent
of the specific methodology used. In all cases the dry merger fraction is almost equal to the wet merger fraction at the lowest-redshift
bin (around $z\sim 0.2$) but the fraction of dry mergers decreases rather quickly between $z=0.2 - 0.3$ and remains somewhat flat at higher redshifts.
In contrast to this, the fraction of wet mergers increases rather rapidly
with redshift, and by $z\sim0.7$ wet mergers outnumber dry ones by a factor of 6 to 1.   The dry merger fraction at $z = 0.5$ is $\sim$ 1\%, which is in agreement with the dry merger fraction of 1\% at $\langle z\rangle \sim 0.55$ obtained by \citet{depropris10}
using close pairs.


Another interesting feature noted in Figure~\ref{merger_fraction} is the way that the total
merger fraction drops from around 6.5\% at $z\sim0.2$ to around $4.5\%$ between
$z=0.3$ and $z=0.5$, before rising to around 7\% at $z=0.7$. This effect seems
due to the rapid growth with time in dry mergers at lower redshifts and the rapid decline with time in wet mergers in earlier Universe.
 This dip is seen more predominantly in D2 than in D1 field suggesting
that cosmic variance is involved \citep{bridge10}. The non-monotonic changes in the merging fraction seen suggest
that over the redshift range being probed a power law provides a poor description
for the changing total merging fraction.
At the same time, the monotonic changes in the wet and dry populations when
considered separately suggests that a power law might provide
a reasonable description for these sub-populations, so we have
analyzed them separately. We adopt
the analytic form $f_M =f_0 \times (1+z)^m$, 
where $f_0$ is also set as a free parameter, and $m$ is the standard power-law index.
Fitting this relationship to the $R/T=0.5$ curves on the right-hand panel yields
 $m=2.0\pm0.3$ and $m=-3.1\pm 0.5$ for wet and
 dry mergers, respectively.  Since the merger sample may not be complete in 
 the last bin at $z=0.77$, and the power-law indices $m=1.2\pm0.4$ and $m=-4.1\pm 0.6$ are obtained
 for wet and
 dry mergers with the last data point excluded.  
 We see that the dry merging population is best described by
 negative evolution, and the basic statement that the wet and dry
 populations are evolving with the same $m$ is excluded
 at more than three sigma significance\footnote{Note that the total and wet merger fraction keep increasing rapidly beyond the redshift limit of this paper, so it seems that the merger rate evolves more dramatically at higher redshifts than at $z < 0.7$.  See Bridge et al. (2010) for more discussion on this point.}.

The careful consideration given to selection effects described in Bridge et al. (2010)
leaves little doubt that the effects noted in Figure~\ref{merger_fraction}
are really seen in the observational data, but what is their
ultimate physical significance?
The visibility of red and blue stellar populations is expected to be a strong function
of k-corrections, which differer markedly for red and blue galaxies at $z\sim0.7$ (the redshift
limit for the present paper). To better understand this, 
we will now explore the space densities of mergers using the
$V_{\rm max}$ formalism to account for color and luminosity biases.

\begin{figure*}[htb]
\begin{center}
\includegraphics[width=7.5cm,angle=0]{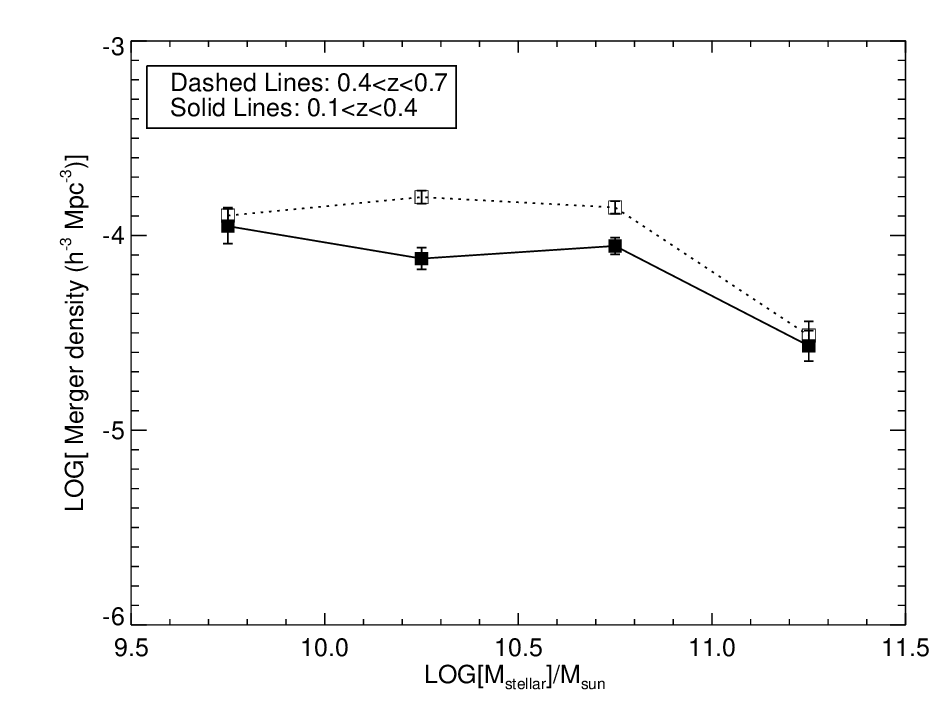}
\includegraphics[width=7.5cm,angle=0]{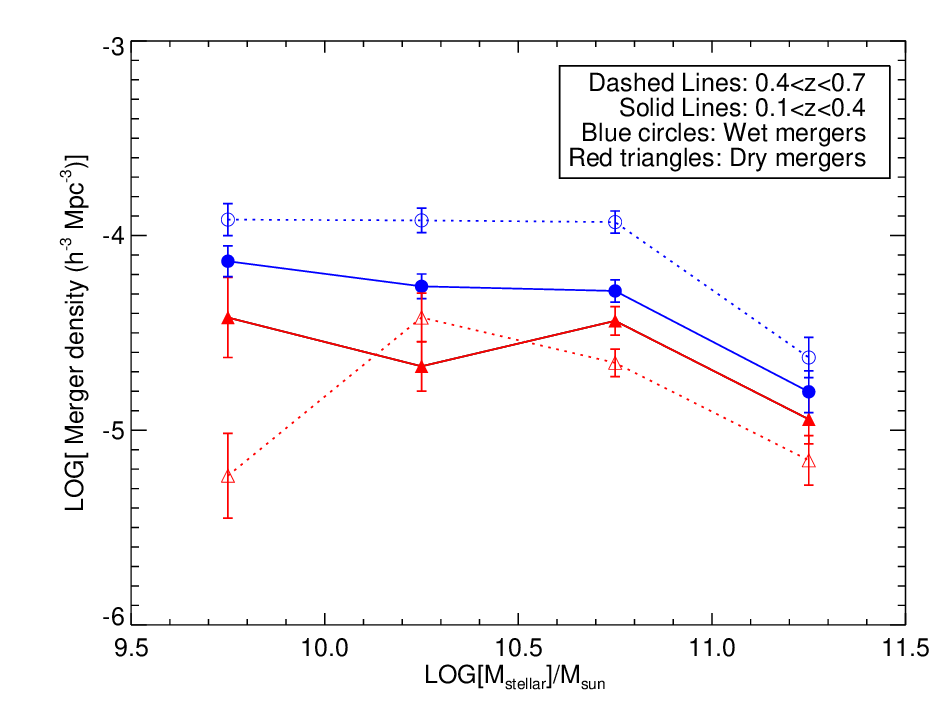}
\end{center}
\caption{[Left:]
Mass density function in mergers in two redshift bins. The high-redshift bin ($0.4 < z <0.7$) is
shown with a dashed curve, while the low redshift bin ($0.1< z < 0.4$) is shown with a solid curve.  We see
evidence for modest evolution in the mass density function, with most change occurring at intermediate
masses, and no change at the high and low mass ends.
[Right:] Subdivision of the data in the left
hand panel into wet and dry mergers. Line styles segregate the data into two redshift bins, as for the
left-hand panel. As in previous figures, red curves correspond to dry mergers, and
blue curves correspond to wet mergers.  For mergers in the most massive bin, an increase in the dry merger space 
density is offset by a decrease in the wet merger space density, so the total space density is nearly conserved. 
At intermediate masses, the mass density function of dry mergers is nearly unchanged in both redshift bins, with perhaps some
evidence for a slight decrease in the space density of intermediate-mass dry mergers at high redshifts. On the other hand,
the mass density function of wet mergers is increasing with redshift.}
\label{massF}
\end{figure*}

\subsection{The Space Density of Merging Galaxies}
\label{sec:number_density}

Figure \ref{number_density} shows the evolving space density of all mergers, as well as of the
sub-samples of wet and dry mergers. These space densities were computed by weighting each merger by
$1/{V_{\rm max}}$ and summing over redshift bins in the standard manner (see, for example, \cite{ty98}). 
As described in \S2.2, $V_{\rm max}$ values were
computed using the Z-PEG code, although it is once again emphasized that in most cases the ultimate
limit on the detectable volume is set by the merger classification limit 
rather than the magnitude detection limit. 
As with the previous figure, the blue and red curve correspond to wet and dry mergers, while the total for the whole
population is shown in black. Error bars were computed by bootstrap resampling.

The main trends noted in our description of Figure~\ref{merger_fraction} remain visible
in Figure \ref{number_density}, most notably the mild negative evolution in the dry mergers
and positive evolution in the wet mergers, with the populations converging at around
$z\sim0.2$. However, a rather striking new feature emerges: {\em the total space density
for all mergers apparently remains almost flat at all redshifts, with a mild increment at $z > 0.5$}. In other words, the decline
with redshift in dry mergers appears almost perfectly offset by a rise with redshift
in the fraction of wet mergers. 



\subsection{The Role of Stellar Mass}

As was noted in the Introduction, stellar mass is the central parameter driving much
of galaxy evolution. What role does stellar mass play in conditioning the effects noted in the previous two figures?
To explore this, we applied a stellar mass cut of 10$^{9.5} \rm{M}_\odot$ and a redshift cut of $z > 0.1$ to our merger sample, leaving
1296 objects in our mass-limited sample. Figure~\ref{massF} shows the stellar mass density function for this total
sample in two redshift bins (left-hand panel), and the result obtained when the total sample is divided
into wet and dry sub-populations (right-hand panel; note we have adopted the same
color scheme as for the previous figures). Error bars were
derived in the same manner as for the merger number density.  

The left-hand panel of Figure~\ref{massF} shows that the increase in the space density of high-redshift mergers 
occurs mostly in intermediate-mass galaxies. There is little evidence for an increase in the space density
of merging galaxies in either our lowest mass bin or in our highest mass bin, which is consistent with the result shown in Figure \ref{number_density}. 
The overall situation is
clarified further in the right hand panel, which shows that: (i) the space
density of wet mergers is always higher than the space density of dry mergers, regardless
of redshift or mass; (ii) the space density of
wet mergers evolves more quickly than the space density of dry mergers. In fact, 
in the dry merging population there appears to be little evidence 
for redshift evolution except in the lowest mass bin, suggesting that any evolutionary trends
for dry mergers occur mainly at low masses.

\begin{figure*}[htb]
\begin{center}
\includegraphics[width=10.5cm,angle=0]{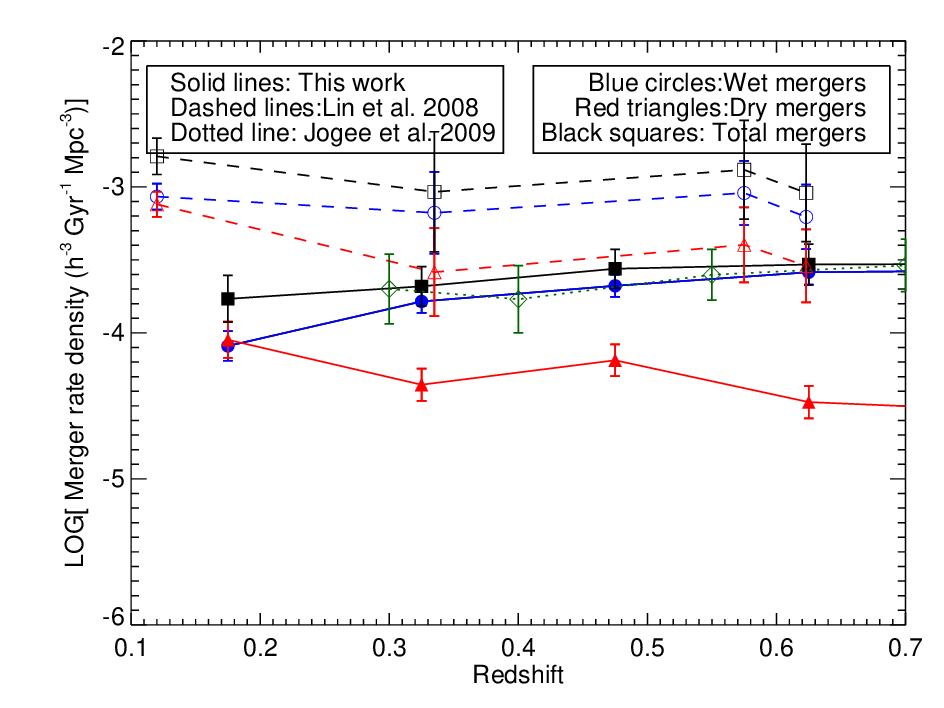}
\end{center}
\caption{A comparison of the evolving merger rate density in the present work with the results obtained by Lin et al. (2008; Lin08) and Jogee et al. (2009; Jog09) using pair counts and automatic morphological selection.  Our merger rate density is derived by choosing a merging time scale of 0.8 $\pm$0.2 Gyr, as discussed in Bridge et al. (2010).  The merging timescale used by Lin08 and Jog09 is 0.5 Gyr.
Qualitative trends determined
using both methods are similar. The absolute merger rate density from our work is lower by a factor of $\sim$3 compared to Lin08
but is comparable to that in Jog09 (see text for details).}
\label{MR_compare}
\end{figure*}

\begin{figure*}[htb]
\begin{center}
\includegraphics[width=10.5cm,angle=0]{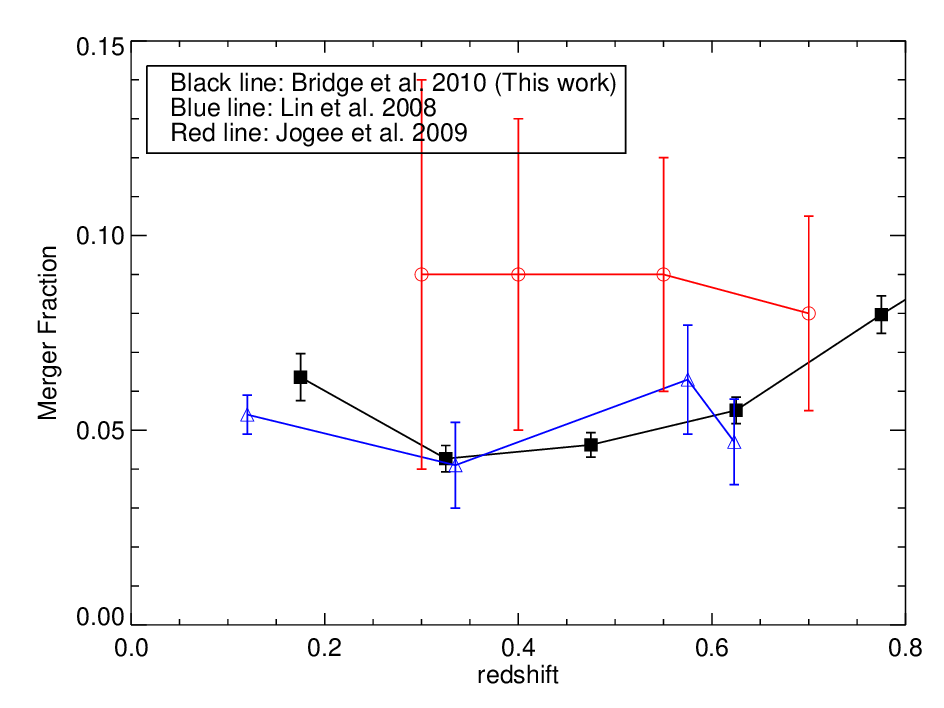}
\end{center}
\caption{Comparison of merger fractions with two different merger selection methods.  The blue line and triangles indicate the merger fraction derived from close pair merger selection method; the red line and circles indicate the merger fraction derived from mergers selected by automated CAS asymmetry and clumpiness parameters.  In Figure \ref{MR_compare} our merger rate density is in good agreement with data obtained by Jog09.  However, it is clear that our merger fraction is in good agreement with data obtained by Lin08.  This indicates that the main uncertainty comes from different merging timescales and correction factors that translate the empirical merger fraction to merger rate density (see text for more details). }
\label{merger_fraction_compare}
\end{figure*}

\begin{figure*}[htb]
\begin{center}
\includegraphics[width=10.5cm,angle=0]{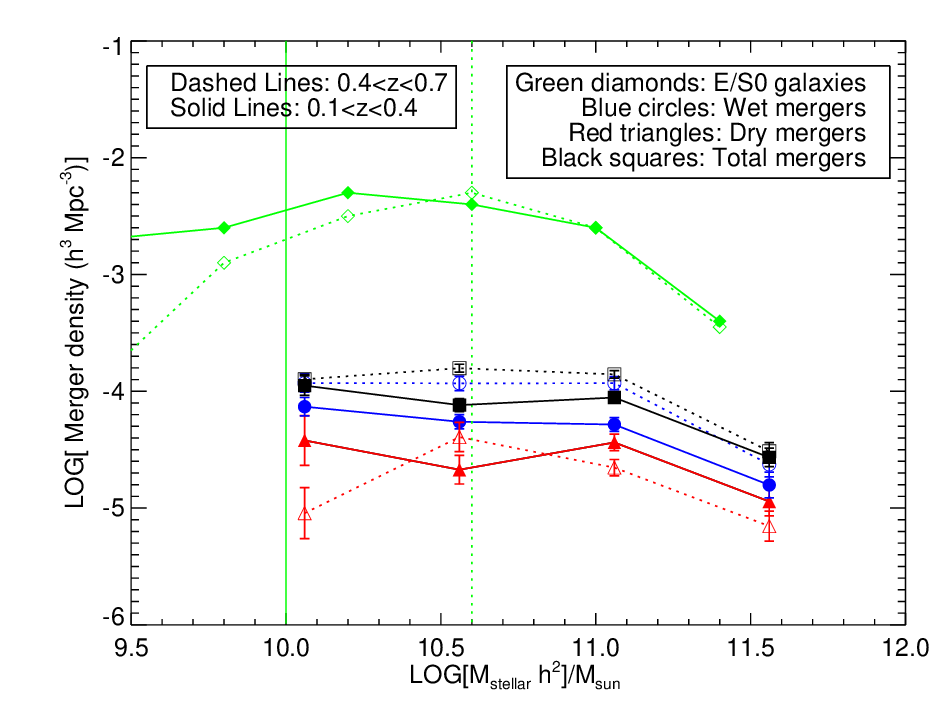}
\end{center}
\caption{A comparison of the stellar mass density function for mergers in the present sample with the corresponding stellar mass density
function for early-type galaxies presented in \citet{bundy05}.  The data point of early type galaxies are more or less from the same redshift bins.  Two vertical lines indicate the completeness limit of early type galaxies in different redshifts. The x-axis is
in units of M$_{\rm stellar}$ h$^2$/M$\rm_{sun}$ to facilitate comparison with \citet{bundy05}.
Note the completely different shapes for these two samples, even though
elliptical galaxies are associated with the end product of mergers, and there is no evolution in the space density of massive
elliptical galaxies. This shows quite clearly that not all of 
the major mergers will end up with massive early-type galaxies, unless the
merger timescale is much longer than usually assumed. See text for details.
}
\label{MD_compare}
\end{figure*}

\section{DISCUSSION}
\label{discussion}

We would like to compare results with those presented in recent papers by Lin et al. (2008), hereafter Lin08, and Jogee et al. (2009), hereafter Jog09.  These authors used different selection methods. Lin08 used close pair selection while Jog09 used morphological criteria quantified using the automated CAS method \citep{conselice00, conselice03_CAS}.  Both sets of authors provide estimates of the volumetric merger rate with respect to redshift, which provides a nice focus for inter-comparisons. The Lin08 merger sample is composed of data from DEEP2 \citep{Deep2}, the Team Keck Treasury Redshift Survey \citep{TKRS}, the Southern Sky Redshift Survey 2 \citep{ssrs2}, the Millennium Galaxy Catalog \citep{MGC, MGC_spec} and the Canadian Network for Observational Cosmology Field Galaxy Redshift Survey 2 \citep{CNOC2}, for a total of 506 close pairs (compared to ~1300 mergers presented in this work).  The Jog09 sample is taken from the GEMS \citep{rix04} with spectrophotometric redshifts and spectral energy distributions taken from COMBO-17 \citep{wolf04}, for a total of $\sim$800 mergers.  

In order to compare our results with Lin08 and Jog09, the merger densities shown in Figure \ref{number_density} must be converted to a merger rate density, and this requires a merging timescale to be assumed.  The appropriate timescale to use is unclear, since it may be a function of a number of factors, such as galaxy mass ratio and gas fraction \citep{lotz09a, lotz09b}.  Additionally, different merger selection methods would select meregrs in different merging stages and therefore require different correction factors as well as merging timescales.  To make sure that we are comparing the same thing in Figure \ref{MR_compare}, we still keep the original merging timescales used in the corresponding papers.  Recent work based on N-body/hydrodynamical simulations for equal mass gas-rich mergers suggests that merging timescales typically range from 0.2 Gyr to 0.9 Gyr, on the basis of a comparison between the tidal features seen and those produced by simulations\footnote{It is worth noting that the most obvious features emerge at the first encounter and the final merging stages \citep{lotz08a}.}.  Note that the simulation results are valid for mergers selected by different methods.  At present we will adopt a merging timescale of $0.8 \pm$0.2 Gyr estimated from visual comparison of both dry and wet mergers presented in Bridge et al. (2010) who also visually examined the snapshots of simulated mergers and the duration in which the galaxies would be classified as interacting' under the criteria used to classify the galaxies presented in the paper.  Figure~\ref{MR_compare} shows the comoving merger rate density for galaxies with redshifts $z<0.7$ in these samples\footnote{Lin08 designate around 30\% of their merger sample as `mixed pairs' (one red galaxy and one blue galaxy). For purposes of comparison with our own data, we have simply included half of the Lin08 galaxies in the wet mergers category and half of them in the dry mergers category. None of the conclusions in this paper are sensitive to this assumption.}. In this figure dashed lines show the results from Lin08, the dotted line corresponds to Jog09, and solid lines show results obtained from our data.  Blue and red colors represent wet and dry mergers, respectively.  Note that Jog09 does not break the sample down into wet and dry mergers, so only the total merger rate density is shown.

Figure~\ref{MR_compare} shows several interesting things. Firstly, we see that the merger rate density from our work is generally a factor of $\sim3$ less than that in Lin08, although it is in excellent agreement with that of Jog09. Note however, that our merger fraction (shown in Figure \ref{merger_fraction_compare}) is actually in rather good agreement with that of Lin08, so that the disagreement between our merger rate density and that of Lin08 is due to differences in translating from an empirically close pair fraction to a merger rate density. This step is more complicated in a pair count analysis than it is in a morphological analysis, because the former involves a number of coefficients (fraction of galaxy pairs to mergers and correction factor for the companion selection effect due to the limited luminosity range) with values that are not well established. The relevant issues are well-described in Lin08, who outline the steps needed to obtain Equation 5 in that paper.   Another factor that may explain the difference with regard to Lin08 include the possibility that we are missing some mergers because their tidal tails have dropped below our surface brightness detection threshold, although this possibility has been investigated in Bridge et al. (2010) and we believe it is not a significant factor (i.e. certainly less than a factor
of two) out to $z=0.7$. 

Another interesting result that emerges from the figure is the fact that in spite of the systematic offset in the absolute rates compared to Lin08, many of the qualitative trends are similar when our results are compared to theirs. In both cases, the density of wet mergers is essentially identical to the density of dry mergers in the nearby Universe. Also in both cases the evolution in the {\em total} merger rate densities at these redshifts is at best modest, and arguably flat. Finally,
in both cases we see that wet mergers and dry mergers have different trends as a function of redshift, and that dry mergers contribute less at higher redshifts than they do at lower redshifts. Since Jog09 do not break their sample into wet and dry mergers, we are unable to compare
such trends with their sample, although as we have already noted our overall merger rate densities are in good agreement with theirs.

It is highly interesting to explore how the trends shown in Figure~\ref{MR_compare} depend on the stellar masses of the mergers. 
This can be most clearly seen by placing the stellar mass density functions for wet and dry mergers (already shown in Figure~\ref{massF}, subdivided
into two redshift bins) into a broader context, namely that which includes the population these
systems are thought to develop into in hierarchical models.  To this end, Figure~\ref{MD_compare} shows 
the mass density function (converted to our chosen IMF) for early-type galaxies at both $z \sim 0.4$ and $z \sim 0.7$ taken from \citet{bundy05},
as well as our derived mass density functions for mergers.  If early-type galaxies are being built up via mergers then the mass density function of early type galaxies is a cumulative quantity that grows with redshift. A simple model that is worth considering in one in which any redshift-dependent trends have no mass dependence. On a logarithmic plot such a uniform multiplicative mass growth simply displaces the mass density function upward, so the normalization of the mass density function changes with redshift, but the shape of the curve stays the same. Although this argument is based on an assumption that the merger products do not participate in subsequent mergers, since the typical merger rate is $\sim 10^{-4}$ Gyr$^{-1}$Mpc$^{-3}$ it is not very likely that a typical early-type galaxy will experience multiple massive mergers within the span of a few billion years. It is clear from Figure~\ref{MD_compare} that {\em the shapes of the mass density function curves are completely different and that of early type galaxies has very little evolution at the high mass end (with observational errors).} This seems rather surprising in light of the expectation that major mergers are thought to be the
progenitor population for early-type galaxies. In that case, one expects that the two curves
would have similar shapes, with the early-type galaxy mass density function
displaced upwards as elliptical galaxies form, unless the merging timescale is
a very strong function of mass. 
We conclude from this analysis that if early-type galaxies are formed from mergers, at some epoch the mass density function of mergers must have been very different from that seen today.

We can explore the implications of the very different forms for the merger and early-type galaxy
mass density functions by adopting the approach first used by \citet{toomre77},
who compared the number of early-type galaxies to the number of merging galaxies
in the Second Reference Catalog \citep{rc2}.
At the high-mass end, we find that the space density of 
massive early-type galaxies is only a factor of five larger than the space density of wet mergers. 
This is a highly puzzling
result to understand in a framework in which massive mergers produce elliptical galaxies, since our
results also show that the merger rate has not changed rapidly since $z$=0.7. Unless the merging timescale is
$\sim 5$ Gyr for massive mergers (about an order of magnitude larger than current estimates) the space
density of massive galaxy mergers would greatly over-produce massive elliptical galaxies if these
mergers all turn into early-type systems.
A better match is seen for elliptical galaxies of intermediate mass ($\sim10^{10.5} M_\odot$), which outnumber
mergers by a factor of about 50 in Figure~\ref{MD_compare}. In this case, a merging timescale of order
$0.5$ Gyr is consistent with mergers producing the space density 
of elliptical galaxies seen in Figure \ref{MD_compare} (again, assuming a
constant merger rate, as is suggested from our data).

We conclude from this analysis that not all massive major mergers (M$_{\rm stellar} > 10^{11}$M$_{\odot}$) will end up being the most massive
early-type galaxies, at least not at $z<0.7$. 
On the other hand, the merging timescales
suggested by numerical simulations
are consistent with the data if we
suppose that less massive
early-type galaxies form via mergers. 
Since low-intermediate mass ellipticals
are 10--100 times more common than their most massive counterparts, the hierarchical
explanation for the origin of early-type galaxies may be correct for the vast majority of
early-types, even if incorrect for the most massive ones. 

If the massive merging systems captured in
Figure~\ref{MD_compare} do not end up as early-type galaxies, what is
their ultimate fate? That is unclear. However, 
 \citet{robertson08,governato09,kannappan09} claim that the
end product of major mergers of gas-rich spiral galaxies is often a
larger spiral galaxy, so that would seem to be a strong
possibility.

\section{CONCLUSION}

We have divided the 1298 merging galaxies in Bridge et al. (2010) into `wet' and `dry' merging categories using two techniques,
one based on integrated colors and the other based on spatially resolved colors. We find that our general results
are independent of the specific method used. Using the $V_{\rm max}$ methodology, we are able to
compute the space density of merging galaxies out to redshift $z\sim0.7$.
The local space density of wet mergers is essentially identical to the local
space density of dry mergers. The space density
of merging galaxies does not change rapidly with redshift, increasing by
less than a factor of two out to $z\sim0.7$.
On the other hand, wet and dry merging
populations show different
evolutionary trends. At higher redshifts 
dry mergers are a smaller fraction of the
total merging galaxy population, while the wet mergers
make a proportionately greater contribution.

We have compared the stellar
mass density function of merging galaxies to the corresponding stellar 
mass density function of early-type galaxies, and find that these
functions are very different in shape, even though the former
are thought to be the progenitor population of the latter.
We also show that the present space density of  massive
galaxy mergers is already similar (within a factor of three) to that
of {\em existing} massive elliptical galaxies. This suggests that not all of the 
massive major mergers will end up with massive early-type galaxies, unless
the merging timescale is much longer than expected.
On the other hand, for systems with masses less than $10^{10.5} M_\odot$, we find that 
 the space density of low-intermediate mass elliptical galaxies is consistent 
 with hierarchical formation
 via mergers. 
 
 
 \acknowledgements
\noindent The authors thank 
NSERC and the Government of Ontario
for funding this research.  We also thank
Damien Le Borgne and Karl Glazebrook
for useful discussions.
 
\bibliography{ms1}
\end{document}